\documentstyle[12pt,subeqn,axodraw]{article}

\font\bit=cmmib10 scaled\magstep1
\def\bphi{\hbox{\bit\char 30}}

\font\bits=cmmib7 scaled\magstep1
\def\bsphi{\hbox{\bits\char 30}}

\newcommand\vek[1]{\mbox{\rmfamily\bfseries\itshape#1}}
\newcommand\vekexp[1]{\mbox{\scriptsize\rmfamily\bfseries\itshape#1}}

\begin{document}

\title{ Exact Solution of a Charge-Asymmetric Two-Dimensional Coulomb Gas }

\author{L. {\v S}amaj$^1$}

\maketitle

\begin{abstract}
The model under consideration is an asymmetric two-dimensional
Coulomb gas of positively $(q_1=+1)$ and negatively $(q_2=-1/2)$ charged
pointlike particles, interacting via a logarithmic potential.
This continuous system is stable against collapse of positive-negative
pairs of charges for the dimensionless coupling constant 
(inverse temperature) $\beta<4$.
The mapping of the Coulomb gas is made onto the complex 
Bullough-Dodd model, and recent results about that integrable
2D field theory are used.
The mapping provides the full thermodynamics (the free energy,
the internal energy, the specific heat) and the large-distance
asymptotics of the particle correlation functions, in the whole
stability regime of the plasma.
The results are checked by a small-$\beta$ expansion
and close to the collapse $\beta=4$ point.
The comparison is made with the exactly solvable symmetric version 
of the model $(q_1=+1, q_2=-1)$, and some fundamental changes in 
statistics caused by the charge asymmetry are pointed out.
 
\end{abstract}


\vfill

\noindent
$^1$ Institute of Physics, Slovak Academy of Sciences, 
D\'ubravsk\'a cesta 9, 842 28 Bratislava, Slovakia;
E-mail: fyzimaes@savba.sk

\newpage

\renewcommand{\theequation}{1.\arabic{equation}}
\setcounter{equation}{0}

\section{Introduction}
A two-dimensional (2D) Coulomb plasma is the continuous system of 
charged particles, plus perhaps a uniformly charged neutralizing
background, confined to a plane and interacting via the logarithmic 
Coulomb potential. 
In this paper, the classical equilibrium statistical mechanics at 
the (dimensionless) inverse temperature $\beta$ is studied.
We will restrict ourselves to the simple point-particle Coulomb
systems: if there are at least two species of charged particles 
with opposite signs, the Coulomb system is stable against the charge
collapse for small enough $\beta<\beta_{\rm col}$ (in 3D, the
collapse point $\beta_{\rm col}\to 0$ and the stability requires
quantum mechanics \cite{Lieb}).
Going beyond the collapse point needs a short-distance (e.g. hard core)
regularization of the Coulomb interaction.
The famous Kosterlitz-Thouless (KT) transition \cite{Kosterlitz} 
of infinite order from a high-temperature conducting phase to a 
low-temperature insulating phase takes place at some 
$\beta_{\rm KT}>\beta_{\rm col}$.
Like in other dimensions, the 2D Coulomb systems admit the
Debye-H\"uckel treatment in the high-temperature limit $\beta\to 0$,
as was rigorously proven in Ref. \cite{Blum}.
As concerns the bulk thermodynamics, based on the scale invariance
of the logarithmic potential, the density derivatives of the free energy, 
like the pressure, can be obtained exactly in the whole stability 
range of temperatures \cite{Salzberg}-\cite{Hauge}.
On the other hand, the temperature derivatives of the free energy,
like the internal energy or the specific heat, are nontrivial
statistical quantities which cannot be obtained by simple means
and they are known only in special cases (see below).

The most studied versions of the Coulomb plasma are the
one-component plasma (OCP) and the symmetric
two-component plasma (TCP), or Coulomb gas.

The 2D OCP of equally, say unit, charged pointlike
particles in a uniform neutralizing background is
formally related to the fractional quantum Hall
effect \cite{Prange,Francesco}.
No collapse occurs.
There are indications that, around $\beta\sim 142$, the fluid system 
undergoes a phase transition to a 2D Wigner crystal \cite{Choquard}.
In a more recent paper \cite{Moore}, the existence of this transition
has been put in doubt.
The model is exactly solvable at $\beta=2$ by mapping onto free fermions,
in the bulk \cite{Jancovici1} as well as in some inhomogeneous
situations (see review \cite{Jancovici2}).

The symmetric 2D TCP consists of oppositely $\pm 1$ charged particles,
no background is present.
The collapse point $\beta_{\rm col}=2$ coincides with 
the exactly solvable free-fermion point of the
equivalent Thirring model \cite{Gaudin}-\cite{Cornu2}.
At this collapse point, for a fixed fugacity, while the free energy
diverges, truncated Ursell functions are nonzero and finite.
Quite recently \cite{Samaj1}, the complete bulk thermodynamics
of the symmetric 2D TCP was derived exactly in the whole
stability region $\beta<2$.
The mapping onto a bulk 2D sine-Gordon theory with a
conformal normalization of the cos-field was made,
and recent results about that field theory were applied.
Subsequently, the surface tension of the same model in contact with
an ideal-conductor \cite{Samaj2} and with an ideal-dielectric
\cite{Samaj3} rectilinear walls was obtained via a mapping
onto integrable 2D boundary sine-Gordon theories with
Dirichlet and Neumann boundary conditions, respectively.
The large-distance behavior of the bulk charge \cite{Samaj4}
and density \cite{Samaj5} correlation 
functions was derived exactly by exploring 
the form-factor theory of the equivalent sine-Gordon model.

Hansen and Viot \cite{Hansen} have introduced a general
asymmetric 2D Coulomb gas which consists of two
species of pointlike particles with positive
and negative charges of arbitrary strengths
$q_{\sigma}$ $(\sigma = 1,2)$.
For $\vert q_1/q_2 \vert$ being an integer, the model describes 
a plasma of electrons and ions of integer valence.
Without any loss of generality, we choose
\begin{equation} \label{1.1}
q_1 = +1 \quad {\rm and} \quad q_2 = -1/Q
\end{equation}
where $Q=1,2,\ldots$ is a positive integer.
The model interpolates between the symmetric TCP
$(Q=1)$ and, after subtracting the kinetic energy of
2-species, the OCP obtained as the extreme asymmetry
case $Q\to \infty$.
The Boltzmann factor of a pair of positive and negative
charges, $r^{-\beta/Q}$, is integrable at short distance
in 2D space for $\beta<2Q$, so the collapse point is
$\beta_{\rm col} = 2Q$ \cite{Hansen}.
In the case of a vanishing but nonzero hard core around
particles, the KT phase transition was conjectured
to take place at $\beta_{\rm KT} = 4Q$ \cite{Samaj6}.
Highly asymmetric Coulomb mixtures in the strong-coupling
regime have attracted much attention in the last years 
\cite{Shklovskii}-\cite{Schmidt} due to the
phenomenon of overcharging (charge inversion), i.e. the situation
when the number of counterions in the vicinity of a macroion
is so high that the macro-charge is overcompensated.

In this work, we report the exact solution of the $Q=2$
asymmetric 2D Coulomb gas, with the lowest degree of charge
asymmetry $q_1=1$ and $q_2=-1/2$, in the whole stability interval
of pointlike particles $\beta<4$.
The exact solution involves the complete bulk thermodynamics 
(the free energy, the internal energy, the specific heat)
and the large-distance behavior of the particle correlation functions.
It is obtained by mapping the underlying Coulomb gas onto a
member of 2D Toda field theories \cite{Mattsson}, namely
the complex Bullough-Dodd (cBD) model \cite{Dodd}-\cite{Mikhailov}
with a conformal normalization of the exponential field.
Recent results about that integrable field theory, derived by using
the Thermodynamic Bethe Ansatz (thermodynamics) and
the form-factor method (correlation functions), are applied.
Since the calculations in the cBD model are based on special
analyticity assumptions, which are not yet 
rigorously proven, we check the results for the plasma by a
small-$\beta$ expansion, using a renormalized Mayer expansion
in density \cite{Deutsch,Jancovici3}, and close to the collapse
$\beta=4$ point, using an electroneutrality sum rule
\cite{Stillinger} combined with an independent-pair conjecture
made by Hauge and Hemmer \cite{Hauge}.
The comparison is made with the symmetric version of the model, 
and the fundamental changes in statistics
due to the charge asymmetry are pointed out.

The paper is organized as follows.
In Section 2, all relevant aspects of the mapping between
the asymmetric $1/-\frac{1}{2}$ 2D Coulomb gas and the cBD model
are presented.
The complete thermodynamics of the Coulomb system is derived
in Section 3.
The asymptotic large-distance behavior of the particle correlation
functions is obtained within the form-factor method in Section 4.
Section 5 is a brief recapitulation with some concluding remarks.
The results are checked by the small-$\beta$ expansions in Appendix A 
and close to the collapse $\beta=4$ point in Appendix B.

\renewcommand{\theequation}{2.\arabic{equation}}
\setcounter{equation}{0}

\section{Field-theoretical representation of the asymmetric 2D Coulomb gas}
We consider the asymmetric TCP made up of two species of
pointlike particles with charges $q_{\sigma}$ $(\sigma=1,2)$
given by (\ref{1.1}).
The particles are confined to an infinite 2D space of points
${\bf r}\in R^2$, and interact with each other by the pair Coulomb
potential.
The Coulomb potential $v$ at spatial position ${\bf r}$, 
induced by a unit charge at the origin, is given by
the 2D Poisson equation
\begin{equation} \label{2.1}
\Delta v({\bf r}) = - 2 \pi \delta({\bf r})
\end{equation}
as follows
\begin{equation} \label{2.2}
v({\bf r}) = - \ln \left( \vert {\bf r}\vert / r_0 \right)
\end{equation}
The length constant $r_0$, which fixes the zero point of the
Coulomb potential, will be set for simplicity to unity.
The interaction energy $E$ of particles $\{ j \}$ reads
\begin{equation} \label{2.3}
E = \sum_{j<k} q_{\sigma_j} q_{\sigma_k}
v(\vert {\bf r}_j - {\bf r}_k \vert)
\end{equation}
Introducing the microscopic density of particles of species
$\sigma$, ${\hat n}_{\sigma}({\bf r}) = \sum_j \delta_{\sigma,\sigma_j}
\delta({\bf r}-{\bf r}_j)$, the microscopic densities of the total
particle number and of the charge are
\begin{equation} \label{2.4}
{\hat n}({\bf r}) = \sum_{\sigma} {\hat n}_{\sigma}({\bf r}),
\quad \quad
{\hat \rho}({\bf r}) = \sum_{\sigma} q_{\sigma} {\hat n}_{\sigma}({\bf r})
\end{equation} 
respectively.
The energy (\ref{2.3}) can be thus written as
\begin{equation} \label{2.5}
E = {1\over 2} \int {\rm d}^2 r \int {\rm d}^2 r'
{\hat \rho}({\bf r}) v(\vert {\bf r}-{\bf r}'\vert)
{\hat \rho}({\bf r}') - {1\over 2} v(0) \sum_j q_{\sigma_j}^2
\end{equation}
where $v(0)$ is the (divergent) self-energy.

We will work in the grand canonical ensemble, with
position-dependent fugacities $z_{\sigma}({\bf r})$
of species $\sigma = 1,2$.
The grand partition function $\Xi$ at inverse temperature $\beta$, 
considered as a functional of the species fugacities, is defined by
\begin{eqnarray} \label{2.6}
\Xi[z_1,z_2] & = & \sum_{N_1=0}^{\infty} \sum_{N_2=0}^{\infty}
{1\over N_1!} {1\over N_2!} \int 
\prod_{j=1}^{N_1} \left[ {\rm d}^2 u_j z_1({\bf u}_j) \right]
\nonumber \\ & & 
\times \prod_{j=1}^{N_2} \left[ {\rm d}^2 v_j z_2({\bf v}_j) \right]
\exp \left[ - \beta E_{N_1,N_2}(u,v) \right] 
\end{eqnarray}
where $E_{N_1,N_2}(u,v)$ denotes the Coulomb interaction energy
(\ref{2.3}), resp. (\ref{2.5}), of $N_1$ particles of type 1
at vector positions $\{ {\bf u}_j \}_{j=1}^{N_1}$ and $N_2$
particles of type 2 at positions $\{ {\bf v}_j \}_{j=1}^{N_2}$.
The statistical quantity $\Xi$ can be expressed in terms 
of a 2D Euclidean theory with the aid of the standard procedure 
(see, e.g., ref. \cite{Minnhagen}).
Using the representation (\ref{2.5}) for $E$ in $\exp(-\beta E)$
and assuming that $-\Delta/(2\pi)$ is the inverse operator of
$v({\bf r})$ [see relation (\ref{2.1})], one applies the
Hubbard-Stratonovich transformation
\begin{equation} \label{2.7}
\exp \left[ - {\beta\over 2} \int {\rm d}^2 r \int {\rm d}^2 r'
{\hat \rho}({\bf r}) v(\vert {\bf r}-{\bf r}'\vert)
{\hat \rho}({\bf r}') \right] =
{\int {\cal D}\phi \exp \left[ \int {\rm d}^2 r
\left( {1\over 16\pi}\phi \Delta \phi + {\rm i}\sqrt{\beta/4}
~ \phi {\hat \rho}\right)\right] \over 
\int {\cal D}\phi \exp \left( \int {\rm d}^2 r
{1\over 16\pi}\phi \Delta \phi\right)}
\end{equation}
where $\phi({\bf r})$ is a real scalar field and 
$\int {\cal D}\phi$ denotes the functional integration
over this field.
The term $\phi \Delta \phi$ can be turned into $- (\nabla \phi)^2$
by using integration by parts.
The summation over $N_1$ and $N_2$ in (\ref{2.6}) then implies
\begin{equation} \label{2.8}
\Xi[z_1,z_2] = {\int {\cal D}\phi \exp \left(
- S[z_1,z_2] \right) \over \int {\cal D}\phi
\exp \left( - S[0,0] \right)}
\end{equation}
where the action of the 2D Euclidean field theory takes the form
\begin{subequations} \label{2.9}
\begin{eqnarray} 
S[z_1,z_2] & = & \int {\rm d}^2 r \left[
{1\over 16\pi} (\nabla \phi)^2 - 
z_1({\bf r}) {\rm e}^{{\rm i}b\phi} - 
z_2({\bf r}) {\rm e}^{-{\rm i}(b/Q)\phi} \right]
\label{2.9a} \\
b^2 & = & \beta/4 \label{2.9b}
\end{eqnarray}
\end{subequations}
Here, the fugacities $z_{\sigma}$ $(\sigma = 1,2)$ are renormalized 
by the self-energy terms $\exp[\beta v(0)q_{\sigma}^2/2]$.
For the homogeneous system with uniform species fugacities,
$z_{\sigma}({\bf r}) = z_{\sigma}$, 
the action (\ref{2.9}) simplifies to
\begin{equation} \label{2.10}
S(z_1,z_2)  =  \int {\rm d}^2 r \left[
{1\over 16\pi} (\nabla \phi)^2 - 
z_1 {\rm e}^{{\rm i}b\phi} - 
z_2 {\rm e}^{-{\rm i}(b/Q)\phi} \right]
\end{equation}
Note that this action is complex, except in the symmetric
case $Q=1$ with $z_1=z_2$.

The field representation of the multi-particle densities
can be obtained from the functional generator $\Xi$,
defined by (\ref{2.8}) and (\ref{2.9}), in a straightforward way.
At the one-particle level, the density of particles of type 
$\sigma (=1,2)$ is given by
\begin{eqnarray} \label{2.11}
n_{\sigma} & = & \langle {\hat n}_{\sigma}({\bf r}) \rangle
\nonumber \\  & = &  z_{\sigma} {1\over \Xi}
{\delta \Xi \over \delta z_{\sigma}({\bf r})}
\big\vert_{\rm uniform}
\end{eqnarray}
Consequently,
\begin{equation} \label{2.12}
n_{\sigma} = z_{\sigma} \langle {\rm e}^{{\rm i}bq_{\sigma}\phi} \rangle
\end{equation}
where $\langle \cdots \rangle$ denotes the averaging over 
the action (\ref{2.10}).
Note that under the shift of the field variable
$\phi\to \phi+\phi_0$ in (\ref{2.10}) (which has no effect on $\Xi$ 
as the whole), the species fugacities
change as $z_1\to z_1 \exp({\rm i}b\phi_0)$ and
$z_2\to z_2 \exp[-{\rm i}(b/Q)\phi_0]$, and therefore
$\Xi$ depends only on the combination $z_1 (z_2)^Q$.
According to the definition (\ref{2.11}), this property
of $\Xi$ implies the neutrality condition
$n_1 - (1/Q) n_2 = 0$.
At the two-particle level, one introduces the two-body densities
\begin{eqnarray} \label{2.13}
n_{\sigma\sigma'}({\bf r},{\bf r}') & = &
\langle {\hat n}_{\sigma}({\bf r}) {\hat n}_{\sigma'}({\bf r}')
\rangle - n_{\sigma} \delta_{\sigma\sigma'} 
\delta({\bf r}-{\bf r}') \nonumber \\ 
& = & z_{\sigma} z_{\sigma'}
{1\over \Xi} {\delta^2 \Xi \over \delta z_{\sigma}({\bf r})
\delta z_{\sigma'}({\bf r}')} \big\vert_{\rm uniform}
\end{eqnarray}
so that
\begin{equation} \label{2.14}
n_{\sigma\sigma'}(\vert {\bf r}-{\bf r}'\vert)
= z_{\sigma} z_{\sigma'} \langle
{\rm e}^{{\rm i} b q_{\sigma} \phi({\bf r})}
{\rm e}^{{\rm i} b q_{\sigma'} \phi({\bf r}')} \rangle
\end{equation}
It is useful to introduce also the pair distribution functions
\begin{equation} \label{2.15}
g_{\sigma\sigma'}({\bf r},{\bf r}') =
\frac{n_{\sigma\sigma'}({\bf r},{\bf r}')}{n_{\sigma}n_{\sigma'}}
\end{equation}
and the (truncated) pair correlation functions
\begin{equation} \label{2.16}
h_{\sigma\sigma'}({\bf r},{\bf r}') =
g_{\sigma\sigma'}({\bf r},{\bf r}') - 1
\end{equation}

In statistical mechanics, the short-distance behavior of 
the pair distribution function is dominated by the Boltzmann 
factor of the pair Coulomb potential 
\cite{Hansen,Jancovici4},
\begin{equation} \label{2.17}
g_{\sigma\sigma'}({\bf r},{\bf r}') \sim
C_{\sigma\sigma'} 
\vert {\bf r}-{\bf r}' \vert^{\beta q_{\sigma}q_{\sigma'}}
\quad {\rm as} \quad \vert {\bf r}-{\bf r}' \vert \to 0
\end{equation}
(provided $\beta$ is small enough as explained below).
The prefactors $C_{\sigma\sigma'}$ are expressible as follows
\begin{equation} \label{2.18}
C_{\sigma\sigma'} = \exp \left[ \beta \left(
\mu_{q_{\sigma}}^{\rm ex} + \mu_{q_{\sigma'}}^{\rm ex}
- \mu_{q_{\sigma}+q_{\sigma'}}^{\rm ex} \right) \right]
\end{equation}
where $\mu_{q_{\sigma}}^{\rm ex}$ is the excess chemical
potential of species $\sigma$ defined by
\begin{equation} \label{2.19}
\beta \mu_{q_{\sigma}}^{\rm ex} = 
\ln \left( {z_{\sigma} \over n_{\sigma}} \right)
\end{equation}
and $\mu_q^{\rm ex}$ with $q$ arbitrary represents an
extended definition of the excess chemical potential
for a ``guest'' particle of charge $q$ put into 
the considered Coulomb gas (\ref{1.1}).
To get the stability region of $\mu_q^{\rm ex}$,
one has to consider the interaction Boltzmann factor of 
the $q$-charge with an opposite charge from the plasma. 
When $q>0$ ($q<0$), the Boltzmann factor with the opposite $-1/Q$
($+1$) plasma charge at distance $r$, $r^{-\beta q/Q}$
($r^{\beta q}$), is integrable at small $r$ if and only if
$\beta<2Q/q$ ($\beta<-2/q$).
$\mu_q^{\rm ex}$ is therefore finite if
$\beta < (Q+1)/\vert q\vert + (Q-1)/q$ and goes to
$-\infty$ outside of this stability region.
According to the formula (\ref{2.18}) this means
that the prefactor $C_{12}$ remains finite in the whole
stability region $\beta<2 Q$, while the prefactors $C_{11}$ and 
$C_{22}$ are finite only if $\beta<Q$.
The divergence of $C_{11}$ and $C_{22}$ in the middle of 
the stability region, at the point $\beta=Q$,  
is accompanied by a change in the short-distance
behavior (\ref{2.17}) of $g_{11}$ and $g_{22}$.
The general analysis in Ref. \cite{Hansen} shows that,
in the short-distance limit $r\to 0$,
\begin{subequations} \label{2.20}
\begin{eqnarray} 
g_{11}(r) & \propto & r^{\beta} , \quad \quad \beta < Q \nonumber \\
& \propto & r^{2 m + \beta [1-m(4Q+1-m)/(2Q^2)]} , \quad \quad
\frac{2 Q^2}{2 Q - m + 1} < \beta < \frac{2 Q^2}{2 Q - m}
\label{2.20a} 
\end{eqnarray}
where $m$ is an integer from the interval $1\le m \le Q$, and
\begin{eqnarray}
g_{22}(r) & \propto & r^{\beta/Q^2} , \quad \quad \beta < Q \nonumber \\
& \propto & r^{2-\beta(2Q-1)/Q^2} , \quad \quad Q < \beta < 2Q
\label{2.20b}
\end{eqnarray}
\end{subequations}
Close to the collapse point $\beta \to 2Q$, one finds
\begin{subequations} \label{2.21}
\begin{eqnarray}
g_{11}(r) & \propto & r^{Q-1} \label{2.21a} \\
g_{22}(r) & \propto & r^{(2/Q)-2} \label{2.21b}
\end{eqnarray}
\end{subequations}
The collapse relation (\ref{2.21b}) tells us that, for $Q>1$,
$g_{22}(r)$ does not vanish as $r\to 0$ (as is intuitively expected 
since equally charged particles repel each other), but diverges.
This is related to a paradoxical clustering of counterions in
the asymmetric plasma.

We now intend to derive the short-distance behavior (\ref{2.17}),
say for $g_{11}$, in terms of the 2D field theory (\ref{2.10}).
According to the definitions (\ref{2.6}), (\ref{2.13}) 
and (\ref{2.15}), we have
\begin{eqnarray} \label{2.22}
g_{11}({\bf r},{\bf r}') & = &
\vert {\bf r}-{\bf r}' \vert^{\beta} \left(
{z_1\over n_1} \right)^2 {1\over \Xi} \sum_{N_1=2}^{\infty}
\sum_{N_2=0}^{\infty} {N_1(N_1-1) \over N_1! N_2!}
\int \prod_{j=1}^{N_1-2} 
\left[ {\rm d}^2 u_j z_1({\bf u}_j) \right]
\prod_{j=1}^{N_2} 
\left[ {\rm d}^2 v_j z_2({\bf v}_j) \right] \nonumber \\
& & \times \exp \left\{ - \beta E_{N_1-2,N_2}(u,v)
-\beta \int {\rm d}^2 r'' \left[ v({\bf r}-{\bf r}'')
+ v({\bf r}'-{\bf r}'') \right] {\hat \rho}({\bf r}'') 
\right\}
\end{eqnarray}
where ${\hat \rho}$ denotes the microscopic charge density
of $N_1-2$ particles of type 1 and $N_2$ particles of type 2.
One shifts $N_1\to N_1-2$ in the summation over $N_1$.
When ${\bf r}' \to {\bf r}$, the exponential in (\ref{2.22})
can be written in the form presented on the lhs of Eq. (\ref{2.7})
with the substitution ${\hat \rho}({\bf r}'') \to
{\hat \rho}({\bf r}'') + 2 \delta({\bf r}''-{\bf r})$.
Using the Hubbard-Stratonovich identity (\ref{2.7}), the previously
generated single term ${\rm i}\sqrt{\beta/4} ~ \phi {\hat \rho}$ 
now involves two terms: ${\rm i}\sqrt{\beta/4} ~ \phi {\hat \rho}
+ {\rm i} 2 \sqrt{\beta/4} \phi({\bf r})$.
One integrates over particle coordinates, except the
fixed ${\bf r}$-coordinate, and ends up with
\begin{subequations} \label{2.23}
\begin{equation} \label{2.23a}
g_{11}(\vert {\bf r}-{\bf r}'\vert)
\sim \vert {\bf r}-{\bf r}'\vert^{\beta} 
\left( {z_1\over n_1} \right)^2
\langle {\rm e}^{{\rm i}2b\phi} \rangle
\quad {\rm as} \quad \vert {\bf r}-{\bf r}'\vert \to 0
\end{equation}
Analogously, one finds that
\begin{eqnarray}
g_{12}(\vert {\bf r}-{\bf r}'\vert) & \sim &
\vert {\bf r}-{\bf r}'\vert^{-\beta/Q} 
\left( {z_1\over n_1} \right) \left( {z_2\over n_2} \right)
\langle {\rm e}^{{\rm i}(1-1/Q)b\phi} \rangle
\quad {\rm as} \quad \vert {\bf r}-{\bf r}'\vert \to 0
\label{2.23b} \\
g_{22}(\vert {\bf r}-{\bf r}'\vert) & \sim & 
\vert {\bf r}-{\bf r}'\vert^{\beta/Q^2} 
\left( {z_2\over n_2} \right)^2
\langle {\rm e}^{-{\rm i}(2b/Q)\phi} \rangle
\quad {\rm as} \quad \vert {\bf r}-{\bf r}'\vert \to 0
\label{2.23c}
\end{eqnarray}
\end{subequations} 
Comparing these relations with Eqs. (\ref{2.17}) - (\ref{2.19}), 
the excess chemical potential of a particle with charge $q$ 
embedded into the considered plasma
is related to the one-point expectation of the
exponential field as follows
\begin{equation} \label{2.24}
\exp \left( - \beta \mu_q^{\rm ex} \right)
= \langle {\rm e}^{{\rm i}q b\phi} \rangle
\end{equation}
Note that relations (\ref{2.12}) are special
cases of the general formula (\ref{2.24}).
According to the analysis after Eq. (\ref{2.19})
and with respect to the relationship (\ref{2.9b}),
the one-point expectation $\langle {\rm e}^{{\rm i}q b\phi} \rangle$
is finite if $4 b^2 < (Q+1)/\vert q\vert + (Q-1)/q$,
and goes to $+\infty$ otherwise.
Comparing relations (\ref{2.23}) with Eqs. (\ref{2.14}) and (\ref{2.15}),
the short-distance behavior of two-point expectations
of the exponential field is given by
\begin{equation} \label{2.25}
\langle {\rm e}^{{\rm i}qb\phi({\bf r})}
{\rm e}^{{\rm i}q'b\phi({\bf r}')} \rangle
\sim \vert {\bf r}-{\bf r}'\vert^{4b^2 q q'}
\langle {\rm e}^{{\rm i}(q+q')b\phi} \rangle
\quad {\rm as} \quad \vert {\bf r}-{\bf r}'\vert \to 0
\end{equation}

In quantum field theory, the model (\ref{2.10}) can be
regarded as the Gaussian conformal field theory
perturbed by the operator $-z_1 \exp({\rm i}b\phi)
-z_2 \exp[-{\rm i}(b/Q)\phi]$.
The species fugacities $z_1$ and $z_2$ are renormalized
by the (divergent) self-energy factors.
To give the parameters $z_1$ and $z_2$ a precise
meaning, one has to fix the normalization of the
adjoint exponential fields.
The normalization which corresponds to the short-distance
limit of the two-point function (\ref{2.25}) is known
as conformal.
Under the conformal normalization, the diverging self-energy
factor disappears from statistical relationships.
This makes the bridge between the underlying asymmetric Coulomb plasma
and the corresponding field theory with the action (\ref{2.10}), 
treated within the Conformal Perturbation theory.

Note that $q$ and $q'$ should be sufficiently small to ensure
that (\ref{2.25}) is the leading short-distance asymptotics.
From the point of view of the above equivalence, the crucial
is the stability of the two-point expectation asymptotics
(\ref{2.25}) for two opposite charges $q=1$ and $q'=-1/Q$.
In that case, the stability region of 
$\langle {\rm e}^{{\rm i}(q+q')b\phi}\rangle$
is $\beta<\beta_{\rm stab} = 2Q^2/(Q-1)$;
for the symmetric Coulomb gas $\beta_{\rm stab}\to \infty$,
for $Q=2$ $\beta_{\rm stab}$ incidently coincides with
$\beta_{\rm KT}=8$ and for any finite $Q$ it holds
$\beta_{\rm stab}>\beta_{\rm col}$.

\renewcommand{\theequation}{3.\arabic{equation}}
\setcounter{equation}{0}

\section{Thermodynamics of an asymmetric Coulomb gas}
There is a large class of integrable 2D field theories,
known as the affine Toda theories, which are based on the 
Dynkin-diagram classification of simple Lie groups 
(for a nice review, see Ref. \cite{Mattsson}).
Let ${\cal G}$ be a simple Lie algebra of rank $r$,
$\{ {\vek e}_1,\cdots,{\vek e}_r \}$ a set of simple
$r$-dimensional roots of the corresponding Dynkin diagram
and $-{\vek e} = \sum_{i=1}^r n_i {\vek e}_i$
(the coefficients $\{ n_i \}$ are called Kac labels)
the maximal root.
The affine Toda theory built on ${\cal G}$ is
defined by the action
\begin{equation} \label{3.1}
S = \int {\rm d}^2 r \left[ \frac{1}{16\pi}
\left( \partial_{\mu} \bphi \right)^2 +
\sum_{i=1}^r z_i {\rm e}^{b {\vekexp e}_i \cdot \bsphi}
+ z_{r+1} {\rm e}^{b {\vekexp e} \cdot \bsphi} \right]
\end{equation}
where the field $\bphi = (\phi_1,\cdots,\phi_r)$ consists of
$r$ real scalar components and $b$ is the real coupling constant.
The fields in Eq. (\ref{3.1}) are normalized so that at
$\{ z_i = 0 \}_{i=1}^{r+1}$
\begin{equation} \label{3.2}
\langle \phi_a({\bf r}) \phi_b({\bf r}') \rangle
= - 4 \delta_{ab} \ln \vert {\bf r}-{\bf r}' \vert
\end{equation}
For the one-component $(r=1)$ case of interest, there exist
two integrable Toda theories.
When ${\cal G} = A_1^{(1)}$ Lie group with $e_1 = 1$ and
$e = -1$, one obtains the sinh-Gordon ($b$ real) or
sine-Gordon ($b$ pure imaginary) models.
The sine-Gordon model is identified with the action (\ref{2.10})
with $Q=1$ and describes the thermodynamics of the symmetric
2D TCP (see Refs. \cite{Samaj1}-\cite{Samaj4}).
When ${\cal G}=A_2^{(2)}$ Lie group with $e_1=1$ and $e=-1/2$,
the action (\ref{3.1}) takes the form
\begin{equation} \label{3.3}
S_{\rm BD} = \int {\rm d}^2 r \left[ \frac{1}{16\pi}
\left( \nabla \phi \right)^2 + z_1 {\rm e}^{b\phi} +
z_2 {\rm e}^{-b\phi/2} \right]
\end{equation}
This theory is known as the Bullough-Dodd (BD) model \cite{Dodd}.
Its complex version, obtained via the substitutions
$b\to {\rm i}b$ and $z_{1,2}\to - z_{1,2}$,
\begin{equation} \label{3.4}
S_{\rm cBD} = \int {\rm d}^2 r \left[ \frac{1}{16\pi}
\left( \nabla \phi \right)^2 - z_1 {\rm e}^{{\rm i}b\phi} -
z_2 {\rm e}^{-{\rm i}(b/2)\phi} \right]
\end{equation}
is referred to as the Zhiber-Mikhailov-Shabat model
\cite{Zhiber,Mikhailov}, or simply the complex 
Bullough-Dodd (cBD) model.
Comparing (\ref{3.4}) with (\ref{2.10}) one observes that
the cBD model with the short-distance normalization
(\ref{2.25}) is the 2D field realization of the $Q=2$ 
asymmetric TCP, i.e. the system of $+1$ and $-1/2$ charged
particles.
According to (\ref{2.9}), the coupling constant $b$ is
related to the inverse temperature by $b^2=\beta/4$ and
the parameters $z_{\sigma}$ $(\sigma=1,2)$ 
represent the renormalized species fugacities.

The particle spectrum of the BD model (\ref{3.3}) consists of
a single neutral particle of mass $m$.
The two-particle $S$-matrix was described in Ref. \cite{Arinshtein}.
The (dimensionless) specific grand potential
\begin{equation} \label{3.5}
- \omega = \lim_{V\to \infty} \frac{1}{V} \ln \Xi
\end{equation}
where $\Xi$ is given by (\ref{2.8}) and $V$ is the volume,
was obtained in terms of the particle mass $m$ in Ref. \cite{Fateev1}
following the Thermodynamic Bethe Ansatz technique
\cite{Zamolodchikov,Fateev2}.
In the same Ref. \cite{Fateev1}, the relation between the particle
mass $m$ and the model parameters $z_{1,2}$ was established under
the conformal normalization, and an explicit formula for the mean
value of the exponential field $\langle {\rm e}^{a\phi}\rangle$
was suggested by exploring a reflection relationship between
the BD model and the 2D Liouville theory \cite{Lukyanov,Fateev3}.
The derivation of the results was based on special analyticity
assumptions, so their verification by various checks is needed.

The spectrum of the cBD model (\ref{3.4}) exhibits an extremely
complicated hierarchy of particles \cite{Smirnov1}.
The fundamental particle is a three-component kink.
The kinks generate one-component bound states (breathers)
and higher kinks, these higher kinks generate new breathers
and new kinks, etc.
The important simplifying fact is that,
in the whole stability interval of interest $0\le b^2<1$
$(0\le \beta <4)$, the lightest particle, the 1-breather,
corresponds to the analytic continuation of the only particle
in the spectrum of the BD model.
Since the lightest particle dominates in the thermodynamic
limit $V\to \infty$, we can apply all results of 
Ref. \cite{Fateev1} presented in the above paragraph,
with the substitutions $b\to {\rm i}b$ and $z_{1,2}\to - z_{1,2}$,
also to the cBD model.
In particular, the specific grand potential 
(\ref{3.5}) takes the form
\begin{subequations} \label{3.6} 
\begin{eqnarray} 
- \omega & = & \frac{m^2}{16 \sqrt{3} \sin(\pi \xi/3) 
\sin(\pi(1+\xi)/3)} \label{3.6a} \\
\xi & = & \frac{b^2}{2-b^2} \label{3.6b}
\end{eqnarray}
\end{subequations}
and the mass of the lightest 1-breather reads
\begin{equation} \label{3.7}
m = \frac{2\sqrt{3} \Gamma(1/3)}{\Gamma(1-\xi/3) \Gamma((1+\xi)/3)}
\left[ \frac{z_1 \pi \Gamma(1-b^2)}{\Gamma(b^2)} \right]^{(1+\xi)/6}
\left[ \frac{2 z_2 \pi \Gamma(1-b^2/4)}{\Gamma(b^2/4)} \right]^{(1+\xi)/3}
\end{equation}
Note that at the collapse point $b^2=1$ $(\beta=4)$ 
the particle mass $m\to\infty$.
The expected divergence of $-\omega$ (\ref{3.6}) at $b^2=1$
is therefore caused by $m^2$, while the prefactor to $m^2$ remains
finite.
Such behavior is opposite to that observed in the symmetric 2D TCP
\cite{Samaj1} at the corresponding collapse point where the mass
of the lightest sine-Gordon breather, $m_1$, is finite and $-\omega$
diverges due to the prefactor to $m_1^2$.
The expectation value of the exponential field is given by
(see also \cite{Baseilhac})
\begin{subequations} \label{3.8}
\begin{eqnarray}
\langle {\rm e}^{{\rm i}a\phi} \rangle & = &
\left[ \frac{z_2}{z_1} 
\frac{2^{-b^2/2}\Gamma(1+b^2)\Gamma(1-b^2/4)}{\Gamma(1-b^2)
\Gamma(1+b^2/4)} \right]^{\frac{2a}{3b}}
\left[ \frac{m \Gamma(1-\xi/3) \Gamma((1+\xi)/3)}{2^{2/3} \sqrt{3}
\Gamma(1/3)} \right]^{2a^2-ab} \nonumber \\
& & \times \exp \left\{ \int_0^{\infty} \frac{{\rm d}t}{t}
\left[ \frac{\sinh((2-b^2)t) \Psi(t,a)}{\sinh(3(2-b^2)t)
\sinh(2t)\sinh(b^2 t)} - 2 a^2 {\rm e}^{-2t} \right] \right\}
\label{3.8a}
\end{eqnarray}
where
\begin{eqnarray}
\Psi(t,a) & = & - \sinh(2abt) \big[ \sinh((4-b^2-2ab)t)
- \sinh((2-2b^2+2ab)t) \nonumber \\
& & \quad \quad \quad \quad \quad
+ \sinh((2-b^2-2ab)t) - \sinh((2-b^2+2ab)t)
\nonumber \\ & & \quad \quad \quad \quad \quad
- \sinh((2+b^2-2ab)t) \big] \label{3.8b}
\end{eqnarray}
\end{subequations}
The integral in (\ref{3.8a}) is well defined if
\begin{equation} \label{3.9}
- \frac{1}{2b} < {\rm Re}(a) < \frac{1}{b}
\end{equation}
Taking into account (\ref{2.12}), the charge-neutrality
condition in the considered Coulomb gas, $n_1 = n_2/2$,
results in the equality
\begin{equation} \label{3.10}
z_1 \langle {\rm e}^{{\rm i}b\phi} \rangle =
\frac{z_2}{2} \langle {\rm e}^{-{\rm i}(b/2)\phi} \rangle
\end{equation}
Using the integral formula for the logarithm of the Gamma
function \cite{Gradshteyn}
\begin{equation} \label{3.11}
\ln \Gamma(z) = \int_0^{\infty} \frac{{\rm d}t}{t} {\rm e}^{-t}
\left[ (z-1) + \frac{{\rm e}^{-(z-1)t}-1}{1-{\rm e}^{-t}} \right] ,
\quad \quad {\rm Re}(z) > 0
\end{equation}
it can be readily verified that the suggested formula (\ref{3.8})
is consistent with the neutrality condition (\ref{3.10}).

We are now ready to derive the basic density-fugacity relationship
for the asymmetric $1/-\frac{1}{2}$ Coulomb gas.
The total particle number density is
\begin{equation} \label{3.12}
n = z_1 \frac{\partial (-\omega)}{\partial z_1} +
z_2 \frac{\partial (-\omega)}{\partial z_2} =
(1+\xi) (-\omega)
\end{equation}
where the auxiliary parameter $\xi$, introduced in (\ref{3.6b}),
is expressible in the inverse temperature $\beta = 4b^2$
as follows
\begin{equation} \label{3.13}
\xi = \frac{\beta}{8-\beta}
\end{equation}
After some algebra, Eqs. (\ref{3.6}) and (\ref{3.7})
give the explicit density-fugacity relationship
\begin{eqnarray} \label{3.14}
\frac{n^{1-\beta/8}}{(z_1 z_2^2)^{1/3}} & = &
\left[ \frac{\sqrt{3}}{4} \frac{\Gamma^2(1/3)}{(1-\beta/8)\pi^2}
\frac{\Gamma(\xi/3)\Gamma((2-\xi)/3)}{\Gamma(1-\xi/3) \Gamma((1+\xi)/3)}
\right]^{1-\beta/8} \nonumber \\ & & \times
\left[ \frac{\pi \Gamma(1-\beta/4)}{\Gamma(\beta/4)} \right]^{1/3}
\left[ \frac{2 \pi \Gamma(1-\beta/16)}{\Gamma(\beta/16)} \right]^{2/3}
\end{eqnarray}
The length constant $r_0$ in (\ref{2.2}) was set to unity.
This can be shown to imply that the fugacity product $z_1 z_2^2$ in 
neutral configurations of (\ref{2.6}) has dimension
$[{\rm length}]^{-6(1-\beta/8)}$, and so Eq. (\ref{3.14}) is
dimensionally correct.
The small-$\beta$ expansion of the rhs of (\ref{3.14}) reads
\begin{equation} \label{3.15}
\frac{n^{1-\beta/8}}{(z_1 z_2^2)^{1/3}} = \frac{3}{2^{2/3}}
\beta^{\beta/8} \exp \left\{ \left[ 2 C + \ln \left( \frac{\pi}{4} \right)
\right] \frac{\beta}{8} + \left[ 3 \psi'(2/3) - 2 \pi^2 \right] 
\frac{\beta^2}{1728} + O(\beta^3) \right\}
\end{equation} 
where $C$ is the Euler number and 
$\psi(x) = {\rm d}[\ln \Gamma(x)]/{\rm d}x$ is the psi function.
The series representation of the first derivative of $\psi$ reads
\begin{equation} \label{3.16}
\psi'(x) = \sum_{j=0}^{\infty} \frac{1}{(x+j)^2}
\end{equation}
The series expansion (\ref{3.15}) is checked up to the indicated order
in Appendix A.1 by using a bond-renormalized Mayer expansion in density
\cite{Deutsch,Jancovici3}.
For a fixed fugacity product $z_1 z_2^2$, relation (\ref{3.14})
implies the expected collapse singularity of the density $n$:
\begin{equation} \label{3.17}
\frac{n}{(z_1z_2^2)^{2/3}} \sim \frac{\sqrt{3}}{2\pi}
\left[ \Gamma\left( \frac{1}{6} \right) 
\Gamma\left( \frac{1}{3} \right) \right]^2 
\left[ \frac{\Gamma(3/4)}{\Gamma(1/4)} \right]^{4/3}
\frac{1}{(1-\beta/4)^{2/3}} \quad \quad {\rm as}\
\beta \to 4^-
\end{equation}
This singularity is reproduced indirectly in Appendix B.1
by applying a (slightly modified) perfect-screening sum rule,
which is another important check of 
the basic result (\ref{3.14}).

To obtain the complete thermodynamics of the asymmetric
$1/-\frac{1}{2}$ Coulomb gas, we pass from the grandcanonical
to the canonical ensemble via the Legendre transformation
\begin{equation} \label{3.18}
F(T;N_1,N_2) = \Omega + \mu_1 N_1 + \mu_2 N_2
\end{equation}
where 
\begin{subequations} \label{3.19}
\begin{eqnarray}
\Omega & = & k_{\rm B} T \omega(\beta,n) V \label{3.19a} \\
- \omega(\beta,n) & = & \left( 1 - \frac{\beta}{8} \right) n \label{3.19b}
\end{eqnarray}
\end{subequations}
$N_1 = N/3$, $N_2 = 2N/3$ with $N=nV$ being the total particle number, and
\begin{equation} \label{3.20}
\mu_{\sigma}(\beta,n) = k_{\rm B} T~ \ln z_{\sigma}(\beta,n) ,
\quad \quad (\sigma = 1, 2)
\end{equation}
is the chemical potential of species $\sigma$.
The dimensionless specific free energy, $f=F/(N k_{\rm B}T)$, is
thence given by
\begin{equation} \label{3.21}
f(\beta,n) = - \left( 1 - \frac{\beta}{8} \right) +
\frac{1}{3} \ln \left( z_1 z_2^2 \right)
\end{equation}
With the aid of the density-fugacity relationship (\ref{3.14}),
one has explicitly
\begin{eqnarray} \label{3.22}
f(\beta,n) & = & - \left( 1 - \frac{\beta}{8} \right) +
\left( 1 - \frac{\beta}{8} \right) \ln \left( \frac{4}{\sqrt{3}} n \right)
+ \left( 1 - \frac{\beta}{4} \right) \ln \pi \nonumber \\
& & - \left( 1 - \frac{\beta}{8} \right) \ln \left[
\frac{\Gamma^2(1/3)}{(1-\beta/8)} \frac{\Gamma(\xi/3) 
\Gamma((2-\xi)/3)}{\Gamma(1-\xi/3) \Gamma((1+\xi)/3)} \right] \nonumber \\
& & - \frac{1}{3} \ln \left[ \frac{\Gamma(1-\beta/4)}{\Gamma(\beta/4)} \right]
- \frac{2}{3} \ln \left[ \frac{2\Gamma(1-\beta/16)}{\Gamma(\beta/16)} \right]
\end{eqnarray}
The (excess) internal energy per particle, $u^{\rm ex} = \langle E \rangle/N$,
and the (excess) specific heat at constant volume per particle, 
$c_V^{\rm ex} = C_V^{\rm ex}/N$, are determined by elementary
thermodynamics as follows
\begin{subequations} \label{3.23}
\begin{eqnarray}
u^{\rm ex} & = & \frac{\partial}{\partial\beta} f(\beta,n) \label{3.23a} \\
\frac{c_V^{\rm ex}}{k_{\rm B}} & = & - \beta^2 
\frac{\partial^2}{\partial \beta^2} f(\beta,n) \label{3.23b}
\end{eqnarray}
\end{subequations}
The specific heat is independent of the particle number density $n$,
which is a specificity of the 2D pointlike Coulomb gases.
The expansion of $c_V^{\rm ex}$ near the collapse point $\beta=4$
results into the Laurent series
\begin{equation} \label{3.24}
\frac{c_V^{\rm ex}}{k_{\rm B}} = \frac{1}{3(1-\beta/4)^2} -
\frac{2}{3(1-\beta/4)} + O(1), \quad \quad \beta \to 4^-
\end{equation}
The leading two terms in (\ref{3.24}) are reproduced indirectly
in Appendix B.2 by applying an independent-pair conjecture
of Hauge and Hemmer \cite{Hauge}.

\renewcommand{\theequation}{4.\arabic{equation}}
\setcounter{equation}{0}

\section{Large-distance behavior of particle correlations}
In a 2D integrable field theory with particle spectrum $\{ \epsilon \}$,
correlation functions of local operators ${\cal O}_a$ ($a$ is a free
parameter) are expressible as infinite convergent series over
multi-particle intermediate states (see e.g. Ref. \cite{Smirnov2}).
For the truncated two-point correlation functions
\begin{equation} \label{4.1}
\langle {\cal O}_a({\bf r}) {\cal O}_{a'}({\bf r}') \rangle_{\rm T}
= \langle {\cal O}_a({\bf r}) {\cal O}_{a'}({\bf r}') \rangle
- \langle {\cal O}_a \rangle \langle {\cal O}_{a'} \rangle
\end{equation}
the series reads
\begin{eqnarray} \label{4.2}
\langle {\cal O}_a({\bf r}) {\cal O}_{a'}({\bf r}') \rangle_{\rm T}
& = & \sum_{N=1}^{\infty} \frac{1}{N!} \sum_{\epsilon_1,\ldots,\epsilon_N}
\int_{-\infty}^{\infty} 
\frac{{\rm d}\theta_1\cdots {\rm d}\theta_N}{(2\pi)^N}
F_a(\theta_1,\ldots,\theta_N)_{\epsilon_1\cdots\epsilon_N} \nonumber \\
& & \times ^{\epsilon_N\cdots\epsilon_1}F_{a'}(\theta_N,\ldots,\theta_1)
\exp \left( - \vert {\bf r}-{\bf r}' \vert \sum_{j=1}^N m_{\epsilon_j}
\cosh \theta_j \right) 
\end{eqnarray}
Here, $\theta\in (-\infty,\infty)$ is the rapidity which parametrizes
the energy $E$ and the momentum $p$ of a particle $\epsilon$ of mass
$m_{\epsilon}$ as follows
\begin{equation} \label{4.3}
E = m_{\epsilon} \cosh \theta , \quad \quad
p = m_{\epsilon} \sinh \theta
\end{equation}
and the normalization constants in the form factors $\{ F_a \}$
depend on the specific form of the operator ${\cal O}_a$.
In what follows, we will consider ${\cal O}_a$ to be an 
exponential field:
\begin{subequations} 
\begin{equation} \label{4.4a}
{\cal O}_a({\bf r}) = \exp \left( a \phi({\bf r}) \right)
\end{equation}
in the case of the BD model with the action (\ref{3.3}), and
\begin{equation} \label{4.4b}
{\cal O}_a({\bf r}) = \exp \left( {\rm i} a \phi({\bf r}) \right)
\end{equation}
\end{subequations}
in the case of the cBD model with the action (\ref{3.4}).

The form-factor representation (\ref{4.2}) is particularly useful
in the large-distance limit $\vert {\bf r}-{\bf r}' \vert \to \infty$.
The dominant contribution in this asymptotic limit comes from
a one-particle intermediate state with the minimum value of
the particle mass $m$, at the point of the vanishing rapidity
$\theta \to 0$.
The consequent exponential decay $\exp(-m\vert {\bf r}-{\bf r}'\vert)$
is multiplied by a slower (inverse power law) decaying function,
whose particular form depends on the one-particle form factors.
For the BD model (\ref{3.3}) with the only particle $(\epsilon = 1)$
in the spectrum, the one-particle form factors $F_a(\theta)_1 = F_a$
and $^1F_{a'}(\theta) = F_{a'}^*$ are presented for the exponential
field (\ref{4.4a}) in Refs. \cite{Acerbi,Brazhnikov}.
The transition to the cBD model (\ref{3.4}) is straightforward
since, as was already mentioned in Section 3,
the analytic continuation of the single particle in the
BD spectrum corresponds to the lightest particle of interest
in the cBD model, the 1-breather with mass $m$ given by (\ref{3.7}).
In particular, after the substitutions $a\to {\rm i}a$ and
$b\to {\rm i}b$ in the one-particle form-factor formulae for
the BD model \cite{Acerbi,Brazhnikov}, the lightest-particle
form factor of the exponential field (\ref{4.4b}) in the
cBD model is given by
\begin{subequations} \label{4.5}
\begin{equation} \label{4.5a} 
\frac{F_a}{\langle {\cal O}_a \rangle} = 4 \rho
\sin \left( \frac{2 \pi a}{3b} \xi \right)
\cos \left( \frac{\pi}{6} \left( 1 + 2\xi - 4\xi\frac{a}{b} \right) \right)
\end{equation}
where
\begin{eqnarray}
\rho & = & {\rm i} \left[ \frac{\sin(\pi/3)}{\sin(2\pi\xi/3)
\sin(2\pi(1+\xi)/3)} \right]^{1/2} \nonumber \\
& & \times \exp \left\{ - 2 \int_0^{\infty} \frac{{\rm d}t}{t}
\frac{\cosh(t/6) \sinh(t\xi/3) \sinh(t(1+\xi)/3)}{\sinh t 
\cosh(t/2)} \right\} \label{4.5b}
\end{eqnarray}
\end{subequations}

According to formulae (\ref{2.12})-(\ref{2.16}) adapted to
the considered asymmetric $1/-\frac{1}{2}$ 2D Coulomb gas,
the pair correlation function $h_{\sigma\sigma'}({\bf r},{\bf r}')$
of species $\sigma$ and $\sigma'$ with the corresponding
charges $q_{\sigma}$ and $q_{\sigma'} \in \{ 1, -1/2 \}$ is 
expressible in terms of the averages over the equivalent
cBD model as follows
\begin{equation} \label{4.6}
h_{\sigma\sigma'}({\bf r},{\bf r}') =
\frac{ \langle {\rm e}^{{\rm i}q_{\sigma}b\phi({\bf r})}
{\rm e}^{{\rm i}q_{\sigma'}b\phi({\bf r}')} \rangle_{\rm T}}{
\langle {\rm e}^{{\rm i}q_{\sigma}b\phi} \rangle
\langle {\rm e}^{{\rm i}q_{\sigma'}b\phi} \rangle}
\end{equation}
Using the form-factor representation (\ref{4.2}) for
${\cal O}_a({\bf r}) = \exp({\rm i}a\phi({\bf r}))$,
the leading contribution to $h_{\sigma\sigma'}({\bf r},{\bf r}')$
in the limit $\vert {\bf r}-{\bf r}'\vert \to \infty$ is
determined by the lightest particle in the cBD model, the
1-breather with mass $m$ given by (\ref{3.7}) and the form
factor $F_a$ given by (\ref{4.5}).
Consequently,
\begin{equation} \label{4.7}
h_{\sigma\sigma'}(r) \sim 
\frac{F_{q_{\sigma}b}}{\langle {\cal O}_{q_{\sigma}b} \rangle}
\frac{F^*_{q_{\sigma'}b}}{\langle {\cal O}_{q_{\sigma'}b} \rangle}
\frac{1}{\pi} \int_{-\infty}^{\infty} \frac{{\rm d}\theta}{2}
{\rm e}^{-m r \cos \theta} \quad \quad {\rm as}\
r \to \infty
\end{equation}
Since
\begin{equation} \label{4.8}
\int_{-\infty}^{\infty} \frac{{\rm d}\theta}{2}
{\rm e}^{-m r \cos \theta} = K_0(m r) \sim 
\left( \frac{\pi}{2 m r} \right)^{1/2} \exp ( - m r)
\end{equation}
at asymptotically large $r$ ($K_0$ is the modified Bessel function
of second kind), we finally arrive at
\begin{equation} \label{4.9}
h_{\sigma\sigma'}(r) \sim - \lambda_{\sigma\sigma'}
\left( \frac{\pi}{2 m r} \right)^{1/2} \exp ( - m r)
\end{equation}
Here,
\begin{eqnarray} \label{4.10}
\lambda_{\sigma\sigma'} & = & \frac{8\sqrt{3}}{\pi}
\exp \left\{ - 4 \int_0^{\infty} \frac{{\rm d}t}{t}
\frac{\cosh(t/6)\sinh(t\xi/3)\sinh(t(1+\xi)/3)}{\sinh t \cosh(t/2)}
\right\}
\nonumber \\
& & \times \frac{1}{\sin(2\pi\xi/3) \sin(2\pi(1+\xi)/3)}
\sin\left( \frac{2\pi}{3}\xi q_{\sigma} \right)
\sin\left( \frac{2\pi}{3}\xi q_{\sigma'} \right)
\nonumber \\ & & \times 
\cos\left( \frac{\pi}{6} \left( 1 +2\xi - 4\xi q_{\sigma} \right) \right)
\cos\left( \frac{\pi}{6} \left( 1 +2\xi - 4\xi q_{\sigma'} \right) \right)
\end{eqnarray}
and the mass $m$ (\ref{3.7}) is expressible in terms of the
inverse Debye length for the considered Coulomb gas
\begin{equation} \label{4.11}
\kappa = (\pi \beta n)^{1/2}
\end{equation} 
by combining Eqs. (\ref{3.6}) and (\ref{3.12}),
\begin{equation} \label{4.12}
m = \kappa \left[ \frac{2\sqrt{3}}{\pi\xi} 
\sin\left( \frac{\pi\xi}{3} \right)
\sin\left( \frac{\pi}{3} (1+\xi) \right) \right]^{1/2} 
\end{equation}
We recall that $\xi = \beta/(8-\beta)$.

In the high-temperature limit, the parameters $\lambda_{\sigma\sigma'}$
(\ref{4.10}) and $m$ (\ref{4.12}) of the large-distance asymptotics
(\ref{4.9}) have the following small-$\beta$ expansions
\begin{subequations}
\begin{eqnarray}
\lambda_{\sigma\sigma'} & = & \beta q_{\sigma} q_{\sigma'}
+ \beta^2 q_{\sigma} q_{\sigma'} \left\{ \frac{1}{24} +
\frac{\pi}{72 \sqrt{3}} \left[ 6(q_{\sigma}+q_{\sigma'})-1\right]
\right\} + O(\beta^3) \label{4.13a} \\
m & = & \kappa \left[ 1 + \frac{\pi \beta}{48 \sqrt{3}} +
O(\beta^2) \right] \label{4.13b}
\end{eqnarray}
\end{subequations}
These small-$\beta$ expansions are checked up to the indicated
order in Appendix A.2 by using the renormalized Mayer expansion.
The leading-order terms 
$\lambda_{\sigma\sigma'} \sim \beta q_{\sigma} q_{\sigma'}$
and $m \sim \kappa$ correspond to the Debye-H\"uckel approximation,
the first corrections to this approximation are implied by the
renormalized Meeron graph.

Let us introduce the charge density and particle number density 
combinations of the pair correlation functions
\begin{subequations}
\begin{eqnarray}
h_{\rho}({\bf r},{\bf r}') & = & \sum_{\sigma,\sigma'} 
n_{\sigma} q_{\sigma} n_{\sigma'} q_{\sigma'}
h_{\sigma\sigma'}({\bf r},{\bf r}') \label{4.14a} \\
h_n({\bf r},{\bf r}') & = & \sum_{\sigma,\sigma'} 
n_{\sigma} n_{\sigma'}
h_{\sigma\sigma'}({\bf r},{\bf r}') \label{4.14b}
\end{eqnarray}
\end{subequations} 
respectively.
It is evident that within the Debye-H\"uckel approximation 
the two-particle correlations are determined at large distances by the
charge-charge correlation function $h_{\rho}$, while
$h_n$ is identically equal to zero.
Taking into account the first correction in $\lambda_{\sigma\sigma'}$
(\ref{4.13a}), the strong division between $h_{\rho}(r)$ 
and $h_n(r)$ disappears:
both of them decay at large $r$ exponentially with the same correlation
length $=1/m$, only the corresponding $\beta$-dependent prefactors
differ from one another.
The prefactor is of the form $\lambda_{\sigma\sigma'} = 
q_{\sigma} q_{\sigma'} \lambda(\beta)$
in the case of the symmetric 2D TCP \cite{Samaj4,Samaj5}, 
and so at any $\beta$ in the stability regime the large-distance
asymptotics of the two-particle correlations are determined exclusively
by $h_{\rho}$ ($h_n$ is related to the heavier 2-breather and
therefore goes to zero faster then $h_{\rho}$).
The asymmetry in the particle charges thus causes a fundamental change
in the relative large-distance behavior of the charge and density 
correlation functions.

Since $\kappa \to \infty$ at the collapse point $\beta=4$, 
the particle mass $m$ (\ref{4.12}) diverges, and $h_{\sigma\sigma'}$
given by (\ref{4.9}) reduces trivially to zero.
On the other hand, the mass of the lightest particle is finite
(for a fixed $z$) at the collapse point for the symmetric 2D TCP
\cite{Samaj4}, and the corresponding Ursell functions
$U_{\sigma\sigma'}({\bf r},{\bf r}') = n_{\sigma} n_{\sigma'}
h_{\sigma\sigma'}({\bf r},{\bf r}')$ have a nontrivial 
large-distance dependence.
Also from this point of view the asymmetry in the particle
charges has a relevant influence on the large-distance
characteristics of particle correlation functions.

\section{Conclusion}
In this paper, we have solved exactly the 2D Coulomb gas of pointlike
charged particles, with the charge asymmetry $q_1=1$ and $q_2=-1/2$, 
via the equivalent 2D cBD field theory.
The previous calculations in the cBD model were based on
special analyticity conjectures.
This is why we check our results for the asymmetric Coulomb gas,
in the small-$\beta$ limit and close to the collapse
$\beta=4$ point.
The small-$\beta$ series expansions are generated using the
renormalized Mayer expansion in density, which is a simplified
Coulomb version of the Feynman diagrammatic perturbation
technique in 2D field theories.
On the other hand, the important check of the results close
to the opposite collapse point, based on screening properties
of the Coulomb plasma, is original and has no counterpart
in the field theory.

The asymmetry in the strength of the particle charges brings two
fundamental modifications in the statistical behavior of the 2D Coulomb
gas in comparison with its symmetric version.
Firstly, the large-distance exponential decays of the charge and
density correlation functions are characterized by the same
correlation length (in the symmetric Coulomb gas \cite{Samaj4,Samaj5},
the charge correlation length is twice larger 
than the density one). 
Such behavior does not occur in the Debye-H\"uckel treatment
of the model.
Secondly, the mass of the lightest particle in the spectrum
of the cBD model, which determines the thermodynamics and 
the asymptotics of the particle correlations, diverges at the
collapse point $\beta=4$.
As a consequence, the truncated particle distributions are
trivially equal to zero at $\beta=4$.
This is in contrast to the symmetric 2D Coulomb gas with finite 
and nonzero particle distributions at its collapse free-fermion point.

The other cases of the asymmetric Coulomb gases do not belong to the
family of integrable 2D Toda field theories.
On the other hand, the 2D OCP, which corresponds to the
extreme charge-asymmetric case, is exactly solvable at its
free-fermion point \cite{Jancovici1,Jancovici2}.
The 2D OCP has a field representation \cite{Brilliantov}
which resembles the one of the 2D complex Liouville model
with a kind of ``background'' charge.
Although this theory does not belong to the Toda theories,
it is integrable at least at the aforementioned free-fermion point.
It might be therefore useful to explore its integrability
properties at an arbitrary temperature, first in the classical limit 
and subsequently at the quantum level. 

\newpage

\renewcommand{\theequation}{A.\arabic{equation}}
\setcounter{equation}{0}

\section*{Appendix A: Small-$\beta$ expansions}

Let us consider a general fluid composed of distinct species
of particles $\{ \sigma \}$ with the corresponding position-dependent
densities $\{ n({\bf r},\sigma) \}$.
The particles $i$ and $j$ interact through the pair potential
$v(i,\sigma_i\vert j,\sigma_j)$, where vector position ${\bf r}_i$
is represented simply by $i$.
The technique of the bond-renormalization in the ordinary Mayer
expansion in density (for details, see Refs. 
\cite{Samaj1,Deutsch,Jancovici3}) is based on an expansion of 
each Mayer function in the inverse temperature $\beta$, and on a
consequent series resummation of two-coordinated field circles.
The renormalized $K$-bonds are given by
$$
{\begin{picture}(32,20)(0,7)
    \Photon(0,10)(32,10){1}{7}
    \BCirc(0,10){2.5} \BCirc(32,10){2.5}
    \Text(0,0)[]{$1,\sigma_1$} \Text(32,0)[]{$2,\sigma_2$}
    \Text(17,23)[]{$K$}
\end{picture}}
\ \ \ = \ \ \
{\begin{picture}(32,20)(0,7)
    \DashLine(0,10)(32,10){5}
    \BCirc(0,10){2.5} \BCirc(32,10){2.5}
    \Text(0,0)[]{$1,\sigma_1$} \Text(32,0)[]{$2,\sigma_2$}
    \Text(17,23)[]{$-\beta v$} 
\end{picture}}\ \ +\ \
{\begin{picture}(64,20)(0,7)
    \DashLine(0,10)(32,10){5}
    \DashLine(32,10)(64,10){5}
    \BCirc(0,10){2.5} \BCirc(64,10){2.5}
    \Vertex(32,10){2.2}
    \Text(0,0)[]{$1,\sigma_1$} \Text(64,0)[]{$2,\sigma_2$}
\end{picture}}\ \ +\ \
{\begin{picture}(96,20)(0,7)
    \DashLine(0,10)(32,10){5}
    \DashLine(32,10)(64,10){5}
    \DashLine(64,10)(96,10){5}
    \BCirc(0,10){2.5} \BCirc(96,10){2.5}
    \Vertex(32,10){2.2} \Vertex(64,10){2.2}
    \Text(0,0)[]{$1,\sigma_1$} \Text(96,0)[]{$2,\sigma_2$}
\end{picture}}\ \ + \ldots
$$
or, algebraically,
\begin{eqnarray} \label{A.1}
K(1,\sigma_1\vert 2,\sigma_2) & = & 
[-\beta v(1,\sigma_1 \vert 2, \sigma_2)] \nonumber \\
& & + \sum_{\sigma_3} \int {\rm d}3~
[-\beta v(1,\sigma_1 \vert 3,\sigma_3)]~
n(3,\sigma_3)~ K(3,\sigma_3 \vert 2,\sigma_2)
\end{eqnarray}
It is straightforward to verify by variation of (\ref{A.1})
that it holds
\begin{equation} \label{A.2}
{\delta K(1,\sigma_1\vert 2,\sigma_2) \over \delta n(3,\sigma_3)}
= K(1,\sigma_1\vert 3,\sigma_3) K(3,\sigma_3\vert 2,\sigma_2)
\end{equation}
The excess Helmholtz free energy $F^{\rm ex}$ is expressible
in the renormalized format as follows
\begin{subequations} \label{A.3}
\begin{equation} \label{A.3a}
- \beta F^{\rm ex}[n] = 
\ \  
\begin{picture}(50,20)(0,7)
    \DashLine(0,10)(40,10){5}
    \Vertex(0,10){2.2} \Vertex(40,10){2.2}
\end{picture}
 + \ \ 
D^{(0)}[n]  +  \sum_{s=1}^{\infty} D^{(s)}[n]
\end{equation}
where
\begin{equation} \label{A.3b}
D^{(0)} = \ \
\begin{picture}(40,20)(0,7)
    \DashCArc(20,-10)(28,45,135){5}
    \DashCArc(20,30)(28,225,315){5}
    \Vertex(0,10){2} \Vertex(40,10){2}
\end{picture} \ \ +\ \ 
\begin{picture}(40,20)(0,19)
    \DashLine(0,10)(40,10){5}
    \DashLine(0,10)(20,37){5}
    \DashLine(20,37)(40,10){5}
    \Vertex(0,10){2} \Vertex(40,10){2} \Vertex(20,37){2}
\end{picture} \ \ +\ \ 
\begin{picture}(30,20)(0,10)
    \DashLine(0,0)(30,0){5}
    \DashLine(0,0)(0,30){5}
    \DashLine(0,30)(30,30){5}
    \DashLine(30,0)(30,30){5}
    \Vertex(0,0){2} \Vertex(30,0){2} \Vertex(0,30){2} \Vertex(30,30){2}
\end{picture} \ \ +\ \ \cdots 
\end{equation}
\vskip 0.7cm
\noindent is the sum of all ring diagrams which cannot undertake the
bond-renormalization procedure and
\begin{equation} \label{A.3c}
\begin{picture}(60,40)(0,7)
    \PhotonArc(20,6)(20,15,165){1}{11}
    \PhotonArc(20,14)(20,195,345){1}{11}
    \Photon(0,10)(40,10){1}{8.5}
    \Vertex(0,10){2} \Vertex(40,10){2}
    \Text(20,-25)[]{$D^{(1)}$}
\end{picture}
\begin{picture}(60,40)(0,7)
    \PhotonArc(20,-10)(28,45,135){1}{9}
    \PhotonArc(20,30)(28,225,315){1}{9}
    \PhotonArc(20,6)(20,15,165){1}{11}
    \PhotonArc(20,14)(20,195,345){1}{11}
    \Vertex(0,10){2} \Vertex(40,10){2}
    \Text(20,-25)[]{$D^{(2)}$}
\end{picture}
\begin{picture}(60,40)(0,16)
    \PhotonArc(32,6)(32,115,170){1}{7.5}
    \PhotonArc(-12,40)(32,295,355){1}{7.5}
    \PhotonArc(9,5)(32,10,70){1}{7.5}
    \PhotonArc(52,40)(32,185,250){1}{7.5}
    \Photon(0,10)(40,10){1}{8}
    \Vertex(0,10){2} \Vertex(40,10){2} \Vertex(20,35){2}
    \Text(20,-15)[]{$D^{(3)}$}
\end{picture}
\SetScale{0.9}
\begin{picture}(60,40)(0,13)
    \Photon(0,0)(0,40){1}{7}
    \Photon(40,0)(40,40){1}{7}
    \Vertex(0,0){2} \Vertex(40,0){2} \Vertex(0,40){2} \Vertex(40,40){2}
    \PhotonArc(20,-20)(28,45,135){1}{8}
    \PhotonArc(20,20)(28,225,315){1}{8}
    \PhotonArc(20,20)(28,45,135){1}{8}
    \PhotonArc(20,60)(28,225,315){1}{8}
    \Text(20,-19)[]{$D^{(4)}$}
\end{picture} 
\begin{picture}(60,40)(0,12)
    \Photon(0,0)(40,0){1}{7}
    \Photon(0,0)(0,40){1}{7}
    \Photon(0,40)(40,40){1}{7}
    \Photon(40,0)(40,40){1}{7}
    \Photon(0,0)(40,40){1}{8}
    \Photon(0,40)(40,0){1}{8}
    \Vertex(0,0){2} \Vertex(40,0){2} \Vertex(0,40){2} \Vertex(40,40){2}
    \Text(20,-19)[]{$D^{(5)}$}
\end{picture}
\SetScale{1}  \  {\rm etc.}
\end{equation}
\vskip 1.5cm
\end{subequations}
\noindent are all remaining completely renormalized graphs.
The free energy is the generating functional for one-body,
two-body, etc. densities.
The density-fugacity relationship is generated via
\begin{subequations} \label{A.4}
\begin{eqnarray}
\ln \left[ \frac{n(1,\sigma_1)}{z(1,\sigma_1)} \right]
& = & \frac{\delta (-\beta F^{\rm ex})}{\delta n(1,\sigma_1)}
\nonumber \\ & = & 
\ \  
\begin{picture}(50,20)(0,7)
    \DashLine(0,10)(40,10){5}
    \BCirc(0,10){2.2} \Vertex(40,10){2.2}
    \Text(0,0)[]{$1,\sigma_1$}
\end{picture}
 + \ \ 
d^{(0)}(1,\sigma_1) + \sum_{s=1}^{\infty} d^{(s)}(1,\sigma_1)
\label{A.4a}
\end{eqnarray}
where $d^{(0)}(1,\sigma_1) = \delta D^{(0)}/\delta n(1,\sigma_1)$
can be readily obtained in the form
\begin{equation} \label{A.4b}
d^{(0)}(1,\sigma_1) = \frac{1}{2!} \lim_{2\to 1}
\left[ K(1,\sigma_1\vert 2,\sigma_2) + 
\beta v(1,\sigma_1\vert 2,\sigma_2) \right] \big\vert_{\sigma_2=\sigma_1}
\end{equation}
and
\begin{equation} \label{A.4c}
d^{(s)}(1,\sigma_1) = \frac{\delta D^{(s)}}{\delta n(1,\sigma_1)}
\end{equation}
\end{subequations}
The direct correlation function is generated via
\begin{subequations} \label{A.5}
\begin{eqnarray} 
c(1,\sigma_1\vert 2,\sigma_2) & = & 
\frac{\delta^2 (-\beta F^{{\rm ex}})}{\delta n(1,\sigma_1) 
\delta n(2,\sigma_2)} \nonumber \\
& = & 
\ \  
\begin{picture}(50,20)(0,7)
    \DashLine(0,10)(40,10){5}
    \BCirc(0,10){2.2} \BCirc(40,10){2.2}
    \Text(0,0)[]{$1,\sigma_1$} \Text(43,0)[]{$2,\sigma_2$}
\end{picture}
\ \  + \ \
c^{(0)}(1,\sigma_1\vert 2,\sigma_2) + \sum_{s=1}^{\infty}
c^{(s)}(1,\sigma_1\vert 2,\sigma_2) \label{A.5a}
\end{eqnarray}
where $c^{(0)}(1,\sigma_1\vert 2,\sigma_2) = \delta^2 D^{(0)}/
\delta n(1,\sigma_1) \delta n(2,\sigma_2)$ corresponds to
the Meeron graph,
\begin{equation} \label{A.5b}
c^{(0)}(1,\sigma_1\vert 2,\sigma_2) = \ \ \
\begin{picture}(50,20)(0,7)
    \PhotonArc(20,-10)(28,45,135){1}{9}
    \PhotonArc(20,30)(28,225,315){1}{9}
    \BCirc(0,10){2.2} \BCirc(40,10){2.2}
    \Text(0,0)[]{$1,\sigma_1$} \Text(43,0)[]{$2,\sigma_2$}
\end{picture} 
= \frac{1}{2!} K^2(1,\sigma_1\vert 2,\sigma_2)
\end{equation}
and
\begin{equation} \label{A.5c}
c^{(s)}(1,\sigma_1\vert 2,\sigma_2) = \frac{\delta^2 D^{(s)}}{\delta 
n(1,\sigma_1) \delta n(2,\sigma_2)}
\end{equation}
\end{subequations}
The pair correlation function $h$, defined by relations
(\ref{2.13}) - (\ref{2.16}), is related to $c$ via the 
Ornstein-Zernike (OZ) equation
\begin{eqnarray} \label{A.6}
h(1,\sigma_1\vert 2,\sigma_2) & = & c(1,\sigma_1\vert 2,\sigma_2)
\nonumber \\
& & + \sum_{\sigma_3} \int {\rm d} 3 ~ c(1,\sigma_1\vert 3,\sigma_3)~
n(3,\sigma_3)~ h(3,\sigma_3\vert 2,\sigma_2)
\end{eqnarray}
Notice that the functional derivatives with respect to the density field 
generate root circles not only at obvious field-circle positions,
but also on $K$ bonds according to formula (\ref{4.2}),
causing their right $K$-$K$ division.

Let us return to the 2D asymmetric Coulomb gas with two kinds of 
particles $\sigma = 1,2$ of charges $(q_1=1, q_2=-1/2)$, interacting via 
the logarithmic interaction
\begin{subequations} \label{A.7}
\begin{eqnarray} 
v(i,\sigma_i\vert j,\sigma_j) & = & q_{\sigma_i} q_{\sigma_j}
v(i,j) \label{A.7a} \\
v(i,j) & = & - \ln \vert i - j \vert \label{A.7b}
\end{eqnarray} 
\end{subequations}
We consider the infinite-volume limit, characterized by
homogeneous densities $n(1,\sigma) = n_{\sigma}$ 
constrained by the neutrality condition 
$\sum_{\sigma} q_{\sigma} n_{\sigma} = 0$,
so that 
\begin{equation} \label{A.8}
n_1 = \frac{n}{3}, \quad \quad n_2 =  \frac{2n}{3}
\end{equation} 
with $n$ being the total particle density.
Two-body functions are both isotropic and translationally invariant, 
$c(1,\sigma_1\vert 2,\sigma_2) = c_{\sigma_1 \sigma_2}(\vert 1-2\vert)$,
$h(1,\sigma_1\vert 2,\sigma_2) = h_{\sigma_1 \sigma_2}(\vert 1-2\vert)$.
It follows from Eq. (\ref{A.1}) that the renormalized bonds 
exhibit the same charge-dependence as the 
interaction under consideration (\ref{A.7}),
\begin{subequations} \label{A.9}
\begin{eqnarray} 
K(1,\sigma_1\vert 2,\sigma_2) & = & q_{\sigma_1} q_{\sigma_2} K(1,2)
 \label{A.9a} \\
K(1,2) & = & - \beta K_0(\kappa \vert 1-2 \vert) \label{A.9b}
\end{eqnarray}
\end{subequations}
Here, $K_0$ is the modified Bessel function of second kind
and
\begin{equation} \label{A.10}
\kappa = \left[ 2 \pi \beta \left( n_1 + \frac{n_2}{4} \right)
\right]^{1/2} = \sqrt{\pi \beta n}
\end{equation}
is the inverse Debye length.

\subsection*{A.1 Density-fugacity relationship}
The small-$x$ expansion of the modified Bessel function
$K_0(x)$ reads \cite{Gradshteyn}
\begin{equation} \label{A.11}
K_0(x) = - \ln \left( \frac{x}{2} \right) \sum_{j=0}^{\infty}
\frac{x^{2j}}{2^{2j}(j!)^2} + \sum_{j=0}^{\infty} 
\frac{x^{2j}}{2^{2j}(j!)^2} \psi(j+1)
\end{equation}
where $\psi(1)=-C$.
Thus, in the uniform regime with the interaction (\ref{A.7})
and the renormalized interaction (\ref{A.9}), Eq. (\ref{A.4b})
yields
\begin{equation} \label{A.12}
d_{\sigma}^{(0)} = \frac{\beta q_{\sigma}^2}{2} \left[
C + \ln \left( \frac{\kappa}{2} \right) \right] ,
\quad \quad \sigma = 1, 2
\end{equation}
The first term on the rhs of (\ref{A.4a}) disappears due to
the charge neutrality.
Consequently,
\begin{equation} \label{A.13}
\ln \left( \frac{n_{\sigma}}{z_{\sigma}} \right)
= \frac{\beta q_{\sigma}^2}{2} \left[ C + \frac{1}{2}
\ln \left( \frac{\pi\beta n}{4} \right) \right]
+ \frac{\partial}{\partial n_{\sigma}} \sum_{s=1}^{\infty} 
\frac{D^{(s)}}{V} , \quad \quad
\sigma = 1, 2
\end{equation}
where $V\to \infty$ is the volume of the system.
For the dimensionless quantity of interest
$n^{1-\beta/8}/(z_1 z_2^2)^{1/3}$, (\ref{A.13}) gives
\begin{equation} \label{A.14}
\frac{n^{1-\beta/8}}{(z_1 z_2^2)^{1/3}} = \frac{3}{2^{2/3}}
\beta^{\beta/8} \exp \left\{ \left[ 2 C + \ln \left( \frac{\pi}{4} \right)
\right] \frac{\beta}{8} - \left( \frac{1}{3} \frac{\partial}{\partial n_1}
+ \frac{2}{3} \frac{\partial}{\partial n_2} \right)
\sum_{s=1}^{\infty} \frac{D^{(s)}}{V} \right\}
\end{equation} 

When all $\{ D^{(s)}/V \}$ are set to zero, we have the Debye-H\"uckel
approximation valid in the $\beta \to 0$ limit and, indeed, (\ref{A.14})
then reproduces the leading term of the exact $\beta$-expansion (\ref{3.15}).
The scaling form of the renormalized interaction bonds (\ref{A.9})
permits us to perform the $\beta$-classification of $D^{(s)}/V$.
Let the given completely renormalized diagram $D^{(s)}$ be composed
of $N_s$ skeleton vertices and $L_s$ bonds.
Every dressed bond $K$ brings the factor $-\beta$ and enforces
the substitution $r'=\kappa r$ which manifests itself as the factor
$1/\kappa^2$ for each field-circle integration $\sim \int r {\rm d}r$.
Since there are $(N_s-1)$ independent field-circle integrations
in $D^{(s)}$ we conclude that
\begin{equation} \label{A.15}
\frac{D^{(s)}(\beta)}{V} = \beta^{L_s-N_s+1} \frac{D^{(s)}(\beta=1)}{V}
\end{equation}
In the sketch (\ref{A.3c}), the only diagram which 
contributes to the $\beta^2$ order is $D^{(1)}$, the next diagrams
$D^{(2)}, D^{(3)}, D^{(4)}$ and $D^{(5)}$ constitute the complete
set of contributions to the $\beta^3$ order, etc.

We are interested in the lowest correction to the Debye-H\"uckel
limit, so only the diagram $D^{(1)}$ has to be analyzed.
The contribution of this diagram is expressible as
\begin{eqnarray} \label{A.16}
\frac{D^{(1)}(n_1,n_2)}{V} & = & \frac{1}{2! 3!} \left(
n_1^2 - 2 n_1 \frac{n_2}{2^3} + \frac{n_2^2}{2^6} \right)
\int {\rm d}^2 r \left[ -\beta K_0(\kappa {\bf r}) \right]^3
\nonumber \\
& = & - \frac{1}{2! 3!} \beta^2 \frac{(n_1-n_2/8)^2}{n_1+n_2/4}
\int \frac{{\rm d}^2 r}{2\pi} K_0^3({\bf r})
\end{eqnarray}
To evaluate the integral in (\ref{A.16}), we can make use of
the 2D Fourier components of $K_0({\bf r})$ and $K_0^2({\bf r})$
(see Appendix of Ref. \cite{Samaj1}) to derive
\begin{equation} \label{A.17}
\int \frac{{\rm d}^2 r}{2\pi} K_0^3({\bf r}) =
\int_0^{\infty} {\rm d}k 
\frac{\ln\left[ (k/2) + \sqrt{1+(k/2)^2}\right]}{(1+k^2)\sqrt{1+(k/2)^2}}
\end{equation}
After the substitution $k=2\sin(t/2)$, one gets
\begin{eqnarray} \label{A.18}
\int \frac{{\rm d}^2 r}{2\pi} K_0^3({\bf r}) & = &
\frac{1}{\sqrt{3}} \sum_{j=1}^{\infty} 
\frac{\sin(\pi j/3)}{j^2} \nonumber \\
& = & \frac{1}{72} \left[ \psi'\left( \frac{1}{6} \right) + 
\psi'\left( \frac{1}{3} \right) - \psi'\left( \frac{2}{3} \right) 
- \psi'\left( \frac{5}{6} \right) \right]
\end{eqnarray}
where $\psi'(x)$ is given by (\ref{3.16}).
Using the readily derivable relationships \cite{Gradshteyn}
\begin{subequations} \label{A.19}
\begin{eqnarray}
\psi'(x) & = & \frac{1}{4} \left[ \psi'\left(\frac{x}{2}\right)
+ \psi'\left(\frac{x+1}{2}\right) \right] \label{A.19a} \\
\psi'(x) & = & \frac{1}{9} \left[ \psi'\left(\frac{x}{3}\right)
+ \psi'\left(\frac{x+1}{3}\right) + \psi'\left(\frac{x+2}{3}\right)
\right] \label{A.19b} \\
\psi'(x) & = & \psi'(x+1) + \frac{1}{x^2} \label{A.19c}
\end{eqnarray}
\end{subequations}
valid for any $x$, one finds
\begin{equation} \label{A.20}
\int \frac{{\rm d}^2 r}{2\pi} K_0^3({\bf r})  =
- \frac{1}{18} \left[ 3 \psi'\left( \frac{2}{3} \right)
- 2 \pi^2 \right]
\end{equation}
The consideration of (\ref{A.20}) in (\ref{A.16}) leads to
\begin{equation} \label{A.21}
- \left( \frac{1}{3} \frac{\partial}{\partial n_1} +
\frac{2}{3} \frac{\partial}{\partial n_2} \right)
\frac{D^{(1)}}{V} = \left[ 3 \psi'\left( \frac{2}{3} \right)
- 2 \pi^2 \right] \frac{\beta^2}{1728}
\end{equation}
where (\ref{A.8}) was taken into account.
Finally, inserting (\ref{A.21}) into (\ref{A.14}),
the term of the order $\beta^2$ is reproduced exactly in the exponential
of the small-$\beta$ expansion (\ref{3.15}), generated from
the exact density-fugacity relationship (\ref{3.14}).

\subsection*{A.2 Large-distance asymptotics of particle correlations}
In the Fourier picture, the OZ equation (\ref{A.6}) reads
\begin{equation} \label{A.22}
{\hat h}_{\sigma\sigma'}(k) = {\hat c}_{\sigma\sigma'}(k)
+ 2\pi \sum_{\sigma''=1,2} {\hat c}_{\sigma\sigma''}(k) n_{\sigma''}
{\hat h}_{\sigma''\sigma'}(k)
\end{equation}
It follows from (\ref{A.5}) that, at the lowest order in $\beta$
(Debye-H\"uckel approximation),
\begin{subequations} \label{A.23}
\begin{eqnarray}
c_{\sigma\sigma'}(r) & =  & \beta q_{\sigma} q_{\sigma'} \ln r
\label{A.23a} \\
{\hat c}_{\sigma\sigma'}(k) & = & - \beta q_{\sigma} q_{\sigma'}
\frac{1}{k^2} \label{A.23b}
\end{eqnarray}
\end{subequations}
Consequently, from (\ref{A.22}),
\begin{subequations} \label{A.24}
\begin{eqnarray}
{\hat h}_{\sigma\sigma'}(k) & = & - \beta q_{\sigma} q_{\sigma'}
\frac{1}{k^2+\kappa^2} \label{A.24a} \\
h_{\sigma\sigma'}(r) & = & - \beta q_{\sigma} q_{\sigma'} K_0(\kappa r)
\label{A.24b} 
\end{eqnarray}
\end{subequations}
and the asymptotic form of $h_{\sigma\sigma'}(r)$ is
\begin{equation} \label{A.25}
h_{\sigma\sigma'}(r) \sim - \beta q_{\sigma} q_{\sigma'}
\left( \frac{\pi}{2\kappa r} \right)^{1/2} \exp(-\kappa r)
\end{equation}
reproducing thus the expected result (\ref{4.9}) with
$\lambda_{\sigma\sigma'} = \beta q_{\sigma} q_{\sigma'}
+ O(\beta^2)$ (\ref{4.13a}) and $m = \kappa [1+O(\beta)]$ (\ref{4.13b}).

With the aid of Eq. (\ref{A.5}), the direct correlation function 
is determined up to the $\beta^2$ order as follows
\begin{equation} \label{A.26}
c_{\sigma\sigma'}(r) = \beta q_{\sigma} q_{\sigma'} \ln r
+ \frac{\beta^2}{2} q_{\sigma}^2 q_{\sigma'}^2 K_0^2(\kappa r)
\end{equation}
Taking $\kappa^{-1} = 1/\sqrt{\pi \beta n}$ as the unit of length,
one has in the Fourier space
\begin{equation} \label{A.27}
{\hat c}_{\sigma\sigma'}(k) = - \beta q_{\sigma} q_{\sigma'}
\frac{1}{k^2} + \frac{\beta^2}{2} q_{\sigma}^2 q_{\sigma'}^2
\frac{\ln \left[ (k/2) + \sqrt{1 + (k/2)^2} \right]}{k~ \sqrt{1+(k/2)^2}}
\end{equation}
The OZ equation (\ref{A.22}) implies
\begin{subequations} \label{A.28}
\begin{eqnarray}
{\hat h}_{11}(k) & = & \frac{1}{d} \left\{ -\beta
+ \frac{\beta^2(3 +4 k^2)}{4 k \sqrt{4+k^2}}
\ln \left[ \frac{k+\sqrt{4+k^2}}{2} \right] \right\} \label{A.28a} \\
{\hat h}_{12}(k) & = & \frac{1}{d} \left\{ \frac{\beta}{2}
+ \frac{\beta^2 k}{4 \sqrt{4+k^2}}
\ln \left[ \frac{k+\sqrt{4+k^2}}{2} \right] \right\} \label{A.28b} \\
{\hat h}_{22}(k) & = & \frac{1}{d} \left\{ -\frac{\beta}{4}
+ \frac{\beta^2(6 + k^2)}{16 k \sqrt{4+k^2}}
\ln \left[ \frac{k+\sqrt{4+k^2}}{2} \right] \right\} \label{A.28c} 
\end{eqnarray}
\end{subequations}
where the denominator is
\begin{equation} \label{A.29}
d = 1 + k^2 - \frac{\beta(2 + 3k^2)}{4 k \sqrt{4+k^2}}
\ln \left[ \frac{k+\sqrt{4+k^2}}{2} \right]
\end{equation}
The asymptotic behavior of $h_{\sigma\sigma'}(r)$ is governed
by the poles of ${\hat h}_{\sigma\sigma'}(k)$ closest to the
real axis.
When $\beta\to 0$, there poles are at $k = \pm {\rm i}$
(or $k^2 = -1$).
In close analogy with Ref. \cite{Samaj4}, to find the $\beta$-correction
of these poles, it is sufficient to expand the $\beta$-dependent
part of the denominator $d$ around say $k={\rm i}$ up to the first
order in $(k-{\rm i})$:
\begin{eqnarray} \label{A.30}
d & = & 1 + k^2 + \frac{\pi \beta}{24 \sqrt{3}} -
\frac{{\rm i}\beta}{108} (9+8\sqrt{3}\pi) (k-{\rm i}) + \cdots
\nonumber \\
& = & (k^2 + m^2) \left[ 1 - \frac{{\rm i}\beta}{108}
\frac{(9+8\sqrt{3}\pi)(k-{\rm i})}{k^2+m^2} + \cdots \right]
\nonumber \\
& = & (k^2 + m^2) \left[ 1 - \beta \left( \frac{1}{24}
+ \frac{\pi}{9\sqrt{3}} \right) + \cdots \right]
\end{eqnarray}
where
\begin{equation} \label{A.31}
m = 1 + \frac{\pi \beta}{48 \sqrt{3}} + O(\beta^2)
\end{equation}
The prefactors $\lambda_{\sigma\sigma'}$, defined by
\begin{equation} \label{A.32}
{\hat h}_{\sigma\sigma'}(k) = - \lambda_{\sigma\sigma'}
\frac{1}{k^2+m^2}
\end{equation}
are deducible from (\ref{A.28}) (with numerators evaluated
at $k^2=-1$) and (\ref{A.30}),
\begin{subequations} \label{A.33}
\begin{eqnarray}
\lambda_{11} & = & \beta + \beta^2 \left( \frac{1}{24}
+ \frac{11\pi}{72\sqrt{3}} \right) + O(\beta^3) \label{A.33a} \\
\lambda_{12} & = & - \frac{\beta}{2} - \frac{\beta^2}{2} 
\left( \frac{1}{24} + \frac{\pi}{36\sqrt{3}} \right) 
+ O(\beta^3) \label{A.33b} \\
\lambda_{22} & = & \frac{\beta}{4} + \frac{\beta^2}{4} 
\left( \frac{1}{24} - \frac{7\pi}{72\sqrt{3}} \right) 
+ O(\beta^3) \label{A.33c}
\end{eqnarray}
\end{subequations}
The large-distance behavior of $h_{\sigma\sigma'}(r)$ is
of type (\ref{4.9}): the prefactors $\lambda_{\sigma\sigma'}$
(\ref{A.33}) and the parameter $m$ (\ref{A.31}) have the
small-$\beta$ expansions (\ref{4.13a}) and (\ref{4.13b}),
respectively, confirming thus the predictions of  
the form-factor method.

\renewcommand{\theequation}{B.\arabic{equation}}
\setcounter{equation}{0}

\section*{Appendix B: Collapse point}
\subsection*{B.1 Density-fugacity relationship}
Based on the analysis in Section 2
applied to our specific $Q=2$ case, close to the
collapse point $\beta=4$ $(b=1)$, the two-body densities
behave like
\begin{subequations} \label{B.1}
\begin{eqnarray}
n_{12}(r) & \sim & z_1 z_2 \langle {\rm e}^{{\rm i}\phi/2}
\rangle \vert_{b\to 1} r^{-\beta/2} \label{B.1a} \\
n_{11}(r) & \propto & r^1 \label{B.1b} \\
n_{22}(r) & \propto & r^{-1} \label{B.1c}
\end{eqnarray}
\end{subequations}
in the short-distance limit $r\to 0$.
The average in (\ref{B.1a}) is taken over the cBD action (\ref{3.4})
with $b\to 1$.

The one-body and two-body densities in a general Coulomb system
satisfy the electroneutrality sum rule \cite{Stillinger}
\begin{equation} \label{B.2}
- q_{\sigma} n_{\sigma} = \sum_{\sigma'} q_{\sigma'}
\int {\rm d}^d r~ n_{\sigma\sigma'}(r)
\end{equation}
where $\sigma$ numerates the charged species.
For our asymmetric $1/-\frac{1}{2}$ 2D Coulomb gas, one has
in particular
\begin{subequations} \label{B.3}
\begin{eqnarray}
- \frac{n}{3} & = & \int {\rm d}^2 r ~ n_{11}(r) -
\frac{1}{2} \int {\rm d}^2 r ~ n_{12}(r) \label{B.3a} \\
\frac{n}{3} & = & \int {\rm d}^2 r ~ n_{12}(r) - 
\frac{1}{2} \int {\rm d}^2 r ~ n_{22}(r) \label{B.3b} 
\end{eqnarray}
\end{subequations}
It follows from (\ref{3.17}) that, for fixed $z_1$ and $z_2$,
the density $n$ exhibits the singularity of type
$(1-\beta/4)^{-2/3}$ as $\beta \to 4^-$.
This singularity originates in (\ref{B.3}) as a result of
the short-distance integration over the two-body densities
$n_{12}(r) \propto r^{-\beta/2}$;
neither $n_{11}\propto r^1$ nor $n_{22}(r)\propto r^{-1}$
give a diverging contribution as $\beta\to 4^-$ after being
integrated out over short distances $r$.
The problem with Eqs. (\ref{B.3}) is that they imply the
dependences of $n$ on the relevant (diverging) integral
$\int {\rm d}^2 r n_{12}(r)$ with two different prefactors.
We have to admit that the collapse mechanism alters the form
of the sum rules (\ref{B.3}).
The particles can be divided into three basic groups:
one third of particles is of type 1 (charge $q_1=+1$
and density $n_1=1/3$), one third of particles is of type 2 
(charge $q_2=-1/2$ and density $n_2=1/3$) and the remaining third 
of particles of type $2'$ (charge $q_{2'}=-1/2$ and density
$n_{2'}=1/3$) is excluded from the collapse phenomenon
(they only feel the charge $+1/2$ of each collapsed pair of 
1,2-particles and do not enter into diverging screening integrals).
Under this assumption, close to $\beta=4$, the electroneutrality sum 
rules (\ref{B.2}) are modified as follows
\begin{equation} \label{B.4}
-q_{2'}n_{2'} - q_{\sigma}n_{\sigma} \sim
q_{\sigma^*} \int {\rm d}^2 r~ n_{\sigma\sigma^*}(r) ,
\quad \quad \sigma= 1, 2
\end{equation}
where $1^*=2$ and $2^*=1$.
The couple of Eqs. (\ref{B.4}) is consistent and implies the
only singular relationship
\begin{equation} \label{B.5}
n \sim 3 \int {\rm d}^2 r~ n_{12}(r) \quad \quad
{\rm as}\ \beta \to 4^-
\end{equation}

We now put the short-distance expansion of $n_{12}(r)$,
Eq. (\ref{B.1a}), into the integral on the rhs of (\ref{B.5}) cut 
at some finite $\vert r\vert =L$ ($L$ is a length over which 
the Coulomb interaction is screened) and obtain
\begin{equation} \label{B.6}
n \sim z_1 z_2 \langle {\rm e}^{{\rm i}\phi/2} \rangle \vert_{b\to 1}
\frac{3\pi}{1-\beta/4} \quad \quad {\rm as}\ \beta \to 4^-
\end{equation}
To evaluate the mean value of the exponential field, we apply
the conjectured formula (\ref{3.8}) for $a=1/2$, $b\to 1$ and
$\xi=1$.
Consequently,
\begin{eqnarray} \label{B.7}
\langle {\rm e}^{{\rm i}\phi/2} \rangle \vert_{b\to 1}
& \sim & \sqrt{2} \left[ \frac{z_2}{z_1} \frac{\Gamma(3/4)}{\Gamma(1/4)}
\left( 1 - \frac{\beta}{4} \right) \right]^{1/3} \nonumber \\
& & \times \exp \left\{ \int_0^{\infty} \frac{{\rm d}t}{2t}
\left[ \frac{1}{\cosh t (2\cosh t-1)} - {\rm e}^{-2t} \right] \right\}
\end{eqnarray}
where we have applied $\Gamma(x+1) = x \Gamma(x)$.
It is simple to show by using the integral representation of the
logarithm of the Gamma function (\ref{3.11}) that
\begin{equation} \label{B.8}
\exp \left\{ \int_0^{\infty} \frac{{\rm d}t}{2t}
\left[ \frac{1}{\cosh t (2\cosh t-1)} - {\rm e}^{-2t} \right] \right\}
= \frac{1}{(2\pi)^2} \left( \frac{2}{3} \right)^{1/2}
\frac{\Gamma(3/4)}{\Gamma(1/4)} \left[ \Gamma\left( \frac{1}{6} \right)
\Gamma\left( \frac{1}{3}\right) \right]^2
\end{equation}
Eqs. (\ref{B.6}) - (\ref{B.8}) reproduce exactly the collapse
singularity (\ref{3.17}) deduced from the exact density-fugacity
relationship (\ref{3.14}).

\subsection*{B.2 Thermodynamics}
Although the thermodynamics of the system close to the collapse
point is derivable directly from the collapse $n-z$ relationship
discussed in the previous subsection, there exists another simpler
derivation in the spirit of an independent-pair approximation
by Hauge and Hemmer \cite{Hauge}.
As was already mentioned, close to the collapse point, from
the total number of $N$ particles, $N/3$ particles of type 1
and $N/3$ particles of type 2 form pairs.
Their statistical weights contribute dominantly to the
configuration integral
\begin{equation} \label{B.9}
Q \propto \left[ \int_0^L {\rm d}^2 r~ r^{-\beta/2}
\right]^{N/3} = \left[ \frac{\pi}{1-\beta/4}
L^{2(1-\beta/4)} \right]^{N/3} \quad \quad {\rm as}\
\beta \to 4^-
\end{equation}
Thence
\begin{equation} \label{B.10}
\frac{c_V^{\rm ex}}{k_{\rm B}} = \beta^2 
\frac{\partial^2}{\partial \beta^2} 
\left( \frac{1}{N} \ln Q \right) 
= \frac{1}{3(1-\beta/4)^2} - \frac{2}{3(1-\beta/4)} + O(1)
\end{equation}
in full agreement with the Laurent series (\ref{3.24}).

\section*{Acknowledgments}
I am indebted to Bernard Jancovici for careful reading
of the manuscript and useful comments.
The support by Grant VEGA 2/7174/20 is acknowledged.

\end{document}